\documentclass[twocolumn,showpacs,preprintnumbers,amsmath,amssymb]{revtex4}

\usepackage{graphicx}
\usepackage{dcolumn}
\usepackage{bm}
\begin{document}

\preprint{APS/123-QED}

\title{Hopf Fibration and Quantum Entanglement in Qubit Systems}

\author{P. A. Pinilla}
 \email{p-pinill@uniandes.edu.co}
\affiliation{Physics Department, Universidad de los Andes, AA4976, Bogot\'a D.C, Colombia}
\author{J. R. Luthra}
 \email{jluthra@uniandes.edu.co}
\affiliation{Physics Department, Universidad de los Andes, AA4976, Bogot\'a D.C, Colombia}

\date{\today}

\begin{abstract}
Based on the geometry of entangled three and two qubit states, we present the connection between the entanglement measure of the three-qubit state defined using the last Hopf fibration and the entanglement measures known as two- and three-tangle. Moreover, the generalization of the geometric representation of four qubit state and a potential entanglement measure is studied using sedenions for the simplification of the Hilbert space $\bm{S}^{31}$ of the four qubit system. An entanglement measure is proposed and the degree of entanglement is calculated for specific states. The difficulties of a possible generalization are discussed.
\end{abstract}


\maketitle

\section{\label{sec:intro}Introduction}

Quantum entanglement can be understood as a physical resource associated with the nonclassical correlations between separate quantum systems. Many current investigations in Quantum Mechanics are directed toward essential questions about the nature and characterization of quantum entanglement. Recently,  entanglement has gained importance in current developments in quantum cryptography and quantum computing. Entanglement measures are well understood for systems of two and three qubits. Nevertheless, entanglement measures for high dimensional systems are still a matter of research, including the challenging topic of multi-qubit systems \cite{CHEN, MOSS, MIY, COFF}. \\

This investigation is focused on a geometrical explanation of entanglement between qubits using Hopf fibration as a helpful mathematical tool for the reduction of the Hilbert space of the composite system. Basically, a Hopf fibration is a map from a higher dimensional unit sphere to a lower dimensional unit sphere which is not null-homotopic. The simplest example of a Hopf fibration is a map from a three-sphere into a two-sphere in three dimensional Euclidean space $\bm{S}^3\stackrel{\bm{S}^1}{\rightarrow}\bm{S}^2$, which helps to define the well known Bloch sphere as the representation of one pure qubit. In this case, two complex numbers are necessary for the normalization condition that depends on four real parameters. These real numbers define a three-sphere and using the Hopf fibration, all the states differing by a global phase are identified with a unique point in the two-sphere \cite{MOSS}. The circle parameterized by a phase is called the ``fibre" and the two-sphere is called the ``base". The Hopf fibration between higher dimensional spheres is useful for the geometrical understanding of two and three qubits and the role of entanglement is very important from this point of view \cite{CHEN, MOSS}.\\

The normalization condition for a pure two qubit system depends on four complex numbers; this means eight real parameters that define a seven-sphere. At this point, it is possible to define two quaternions such that this normalization will be completely analogous to the case of a single qubit. The second Hopf fibration maps a seven-sphere to a four-sphere; the fibre is now a three-sphere and the base is a four-sphere $\bm{S}^7\stackrel{\bm{S}^3}{\rightarrow}\bm{S}^4$,  this means that all the states differing by a unit quaternion are mapped onto the same value in the base.  Usually, this fibration reduces the two original quaternions to one, but in the separable situation this quaternion simplifies to a complex number and the base is reduced to an ordinary sphere, that is, the Bloch sphere of one of the qubits.  Therefore, the Hopf fibration can determine if the two-qubit state is entangled or separable \cite{CHEN, MOSS}.\\

The next step in this iteration is the quantum system of three qubits, in which a visible difference appears in this instance because the separability among the qubits can be possible in two different ways: First, the three qubits can be separated in the subspace of a single qubit and the subspace of two qubits (bi-separable); and second when the three qubits are fully separated. With the third fibration, the Hilbert space of the three-qubit system, a fifteen-sphere is mapped onto an eight-sphere as base and a seven-sphere as fibre $\bm{S}^{15}\stackrel{\bm{S}^7}{\rightarrow}\bm{S}^8$; this fibration is also entanglement sensitive \cite{CHEN} since in the bi-separable  case the fibration maps once more into the subspace of pure complex numbers. The fully separated case is acquired from the second Hopf fibration acting over the fibre which can be reduced to the multiplication of two Bloch spheres.\\

The goal of this paper is the generalization of this idea for four-qubit systems and an entanglement measure is proposed and  tested for well-known states.  The paper is organized as fallow: In section 2 we recall the work and results of \cite{CHEN, MOSS} and we make the identification between the entanglement measure introduced in \cite{CHEN} and  the two and three-tangle \cite{MIY, COFF}. In section 3 we give a generalization of the Hopf fibration idea for four-qubit systems and an entanglement measure is introduced and compared  with the existing measures in literature for some specific states. Finally, a conclusion about a possible generalization for the n-qubit case is discussed.

\section{\label{sec:sec2}Geometric representation of two and three qubits}

\subsection{\label{sec:sbusec21}Two qubits and second Hopf fibration}
The simplest case of entanglement in Quantum Mechanics is the system made up of two qubits in which  the entanglement measures are well explained. A composite two qubit pure system reads
\begin{equation}
\label{eq:twoqubits}
\left|\Psi\right\rangle=a_{00}\left|00\right\rangle+a_{01}\left|01\right\rangle+a_{10}\left|10\right\rangle+ a_{11}\left|11\right\rangle
\end{equation}
with
\begin{equation}
\label{eq:normalization}
a_{00}, a_{01}, a_{10}, a_{11} \in \mathbb{C}, \qquad \left|a_{00}\right|^2+\left|a_{01}\right|^2+\left|a_{10}\right|^2+\left|a_{11}\right|^2=1
\end{equation}
and the state (\ref{eq:twoqubits}) is separable if and only if
\begin{equation}
\label{eq:separability}
a_{00}a_{11}=a_{01}a_{10}.
\end{equation}
The normalization condition (\ref{eq:normalization}) identifies the Hilbert space of two qubits to a seven-dimensional sphere $\bm{S}^7$, embedded in $\mathbb{R}^8$. The Hopf fibration formulation is a composition of two maps \cite{CHEN},
\begin{subequations}
\label{eq:wholemap}
\begin{eqnarray}
	h_1:\qquad\bm{S}^7 &\rightarrow& \mathbb{R}^4+\left\{\infty\right\}\nonumber\\
			\mathbb{Q}\otimes \mathbb{Q} &\rightarrow& \mathbb{Q}\cup\left\{\infty\right\} \approx \bm{S}^4\nonumber\\
  		\left(q_1, q_2\right)&\rightarrow& h_1= q_1q_2^{-1}	\qquad q_1, q_2 \in \mathbb{Q}
\label{subeq:h1}
\end{eqnarray}
\begin{eqnarray}
	h_2: \qquad\mathbb{R}^4+\left\{\infty\right\} &\rightarrow& \bm{S}^4\nonumber\\
			 \mathbb{Q}\cup\left\{\infty\right\} &\rightarrow& \bm{S}^4
\label{subeq:h2}
\end{eqnarray}
\begin{equation}
				h_2\circ h_1(q_1,q_2)=X_i=\left\langle \sigma_i\right\rangle\qquad i={1,2,3,4,5}
\label{subeq:h1h2}
\end{equation}
\end{subequations}
where $\sigma_i$ in (\ref{subeq:h1h2}) are the Pauli matrices in quaternionic space. The quaternions used in (\ref{subeq:h1}) are defined from the complex coefficients of the state $\left|\Psi\right\rangle$ in (\ref{eq:twoqubits}) as
\begin{equation}
\label{eq:quaternions}
q_1=a_{00}+a_{01}\bm{i}_2 \qquad q_2=a_{10}+a_{11}\bm{i}_2 
\end{equation}
such that the first part (\ref{subeq:h1}) of the whole map gives,
\begin{equation}
\label{eq:first}
h_1=\frac{\left(\bar a_{00} a_{10}+ \bar a_{01} a_{11}\right)+\left(a_{00} a_{11}-a_{01}a_{10}\right)\bm i_2}{\sqrt{\left|a_{10}\right|^2+\left|a_{11}\right|^2}}.
\end{equation}
The value of the quaternion $h_1$ is entanglement sensitive with condition (\ref{eq:separability}) since in the separable case $h_1$ is now a simple complex number. The second map $h_2$ (\ref{subeq:h2}) is an inverse stereographic projection and it sends any two qubit state into points on $\bm{S}^4$ with  coordinates $(X_1, X_2, X_3, X_4, X_5)$ such that $\sum_{i=1}^{i=5}X_i^2=1$. The inverse stereographic projection from the north pole $(1,0,0,0,0)$ onto the equatorial plane leave, as a result, the following coordinates:
\begin{eqnarray}
\label{coordinates}
X_1&=&\left|q_1\right|^2-\left|q_2\right|^2\nonumber\\
X_2&=&2\mathrm{Re}\left(\bar a_{00}a_{10}+\bar a_{01}a_{11}\right)\nonumber\\
X_3&=&2\mathrm{Im}\left(\bar a_{00}a_{10}+\bar a_{01}a_{11}\right)\nonumber\\
X_4&=&2\mathrm{Re}\left(a_{00}a_{11}-a_{01}a_{10}\right)\nonumber\\
X_5&=&2\mathrm{Im}\left(a_{00}a_{11}-a_{01}a_{10}\right)
\end{eqnarray} 
When the state $\left|\Psi\right\rangle$ is separable, the coordinates $X_4$ and $X_5$ are immediately zero. Otherwise, these two coordinates have the information about the entanglement and they are  directly related with the concurrence defined by Wootters \cite{WOOT}, since $X_4^2+X_5^2$ is the concurrence squared. In this situation, the information about the first qubit and its entanglement with the second qubit is stored in the base space $\bm S^4$ and the information about the second qubit is in the fibre space $\bm S^3$. For separable states, the original Hilbert space $\bm S^7$ simplifies to $S^2\times S^2$. 


\subsection{\label{sec:sbusec22}Three qubits and third Hopf fibration}

A general composite three qubit pure system is given by
\begin{eqnarray}
\label{eq:threequbits}
\left|\Psi\right\rangle&=&a_{000}\left|000\right\rangle+a_{001}\left|001\right\rangle+a_{010}\left|010\right\rangle+a_{011}\left|011\right\rangle\nonumber\\
&&+a_{100}\left|100\right\rangle+a_{101}\left|101\right\rangle+a_{110}\left|110\right\rangle+a_{111}\left|111\right\rangle;\nonumber\\
&a_{000}&, a_{001}, a_{010}, a_{011}, a_{100}, a_{101}, a_{110}, a_{111} \in \mathbb{C}
\end{eqnarray}
in this case, the normalization condition reads,
\begin{eqnarray}
\label{eq:normalization2}
&&\left|a_{000}\right|^2+\left|a_{001}\right|^2+\left|a_{010}\right|^2+\left|a_{011}\right|^2+\nonumber\\
&&\left|a_{100}\right|^2+\left|a_{101}\right|^2+\left|a_{110}\right|^2+\left|a_{111}\right|^2=1
\end{eqnarray}
A visible difference appears in this case in which the separability among the qubits can be possible by two different ways: First, the three qubits can be separated in the subspace of a single qubit with basis $\left\{\left|0\right\rangle,\left|1\right\rangle\right\}$ and the subspace of the other two qubits $\left\{\left|00\right\rangle,\left|01\right\rangle,\left|10\right\rangle,\left|11\right\rangle\right\}$
\begin{eqnarray}
\label{eq:biseparable}
\left|\Psi\right\rangle &=&\left(a\left|0\right\rangle+b\left|1\right\rangle\right)\otimes\nonumber\\ &&\left(c\left|00\right\rangle+d\left|01\right\rangle+e\left|10\right\rangle+f\left|11\right\rangle\right)
\end{eqnarray}
and consequently the separability conditions are:
\begin{eqnarray}
\label{eq:conditions2}
a_{000}a_{101}&=&a_{110}a_{011} \qquad a_{000}a_{100}=a_{110}a_{010}\nonumber\\ 
a_{000}a_{111}&=&a_{110}a_{001} \qquad a_{001}a_{101}=a_{111}a_{011}\nonumber\\
a_{001}a_{100}&=&a_{111}a_{010} \qquad a_{010}a_{101}=a_{100}a_{011}.
\end{eqnarray}
For the case when the three qubits are fully separated, two steps are needed to explain entanglement: one in which the three qubits are separable by (\ref{eq:biseparable}), followed by the condition that the two qubit subset is separable in each qubit subset and the procedure is as in section \ref{sec:sbusec21}. In this situation, the normalization condition (\ref{eq:normalization2}) identifies the Hilbert space of three qubits to a fifteen-dimensional sphere $\bm{S}^{15}$, embedded in $\mathbb{R}^{16}$. The last Hopf fibration formulation is a composition of the following two maps \cite{CHEN},
\begin{subequations}
\label{eq:wholemap2}
\begin{eqnarray}
	h'_1:\qquad\bm{S}^{15} &\rightarrow& \mathbb{R}^8+\left\{\infty\right\}\nonumber\\
			\mathbb{O}\otimes \mathbb{O} &\rightarrow& \mathbb{O}\cup\left\{\infty\right\} \approx \bm{S}^8\nonumber\\
  		\left(o_1, o_2\right)&\rightarrow& h'_1= o_1o_2^{-1}	\qquad o_1, o_2 \in \mathbb{O}
\label{subeq:h11}
\end{eqnarray}
\begin{eqnarray}
	h'_2: \qquad\mathbb{R}^8+\left\{\infty\right\} &\rightarrow& \bm{S}^8\nonumber\\
			 \mathbb{O}\cup\left\{\infty\right\} &\rightarrow& \bm{S}^8
\label{subeq:h22}
\end{eqnarray}
\begin{equation}
				h'_2\circ h'_1(o_1,o_2)=X_i=\left\langle \sigma_i\right\rangle\qquad i={1,...,9}
\label{subeq:h11h22}
\end{equation}
\end{subequations}
where $\left\langle \sigma_i\right\rangle$ in (\ref{subeq:h11h22}) are given by
\begin{eqnarray}
\label{eq:Opauli}
\sigma_1&=&\left(
\begin{array}{ll}
0 & 1 \\
1 & 0
\end{array}
\right)\nonumber\\
\sigma_{2,3,4,5,6,7,8}&=&\left(
\begin{array}{ll}
0                   & i_{1,2,3,4,5,6,7} \\
-i_{1,2,3,4,5,6,7}  & 0
\end{array}
\right)\nonumber\\
\sigma_9&=&\left(
\begin{array}{ll}
1    & 0 \\
0    & -1
\end{array}
\right)
\end{eqnarray}
and the octonions in (\ref{subeq:h11}) are defined from the complex coefficients of the state $\left|\Psi\right\rangle$ in (\ref{eq:threequbits}). From the Cayley-Dickson construction:
\begin{eqnarray}
\label{eq:CDconstruction}
q_1=a_{000}+a_{001}\bm{i}_2\qquad q_2&=&a_{010}+\bar a_{011}\bm{i}_2\nonumber\\
q_3=a_{100}+a_{101}\bm{i}_2\qquad q_4&=&a_{110}+\bar a_{111}\bm{i}_2\nonumber\\
o_1=q_1+q_2\bm{i}_4\qquad o_2&=&q_3+q_4\bm{i}_4
\end{eqnarray}
and the first part of the map (\ref{subeq:h11}) is \cite{CHEN}
\begin{eqnarray}
\label{eq:first2}		
h'_1(o_1,o_2)&=&o_1o_2^{-1}\nonumber\\
&=&\frac{K_1+K_2\bm{i}_2+K_3\bm{i}_4+K_4\bm{i}_6}{\left|a_{100}\right|^2+\left|a_{101}\right|^2+\left|a_{110}\right|^2+\left|a_{111}\right|^2}
\end{eqnarray}
\begin{eqnarray}
\label{eq:constants}	
K_1&=&a_{000}\bar a_{100}+a_{001}\bar a_{101}+\bar a_{110}a_{010}+\bar a_{111}a_{011}\nonumber\\
K_2&=&a_{001}a_{100}-a_{000}a_{101}+\overline{\left(a_{110}a_{011}-a_{111}a_{010}\right)}\nonumber\\
K_3&=&a_{010}a_{100}-a_{110}a_{000}+\overline{\left(a_{111}a_{001}-a_{011}a_{101}\right)}\nonumber\\
K_4&=&a_{110}a_{001}-a_{010}a_{101}+\overline{\left(a_{111}a_{000}-a_{011}a_{100}\right)}.\nonumber\\
&&
\end{eqnarray}
$h'_1$ is also sensitive to the conditions in (\ref{eq:conditions2}), and $h'_1$ maps into the subspace of pure complex numbers $\mathbb{C}\cup\infty$ in the octonionic field $\textbf{O}\cup\infty$ \cite{CHEN}. As discussed in the previous section, the  coordinates $X_3, X_4, X_5, X_6, X_7, X_8$ are zero for separable states, and non-null for entangled states. These coordinates characterize the degree of  entanglement of one qubit with the other two qubits and it is possible to define a measure $E$ as \cite{CHEN}
\begin{equation}
	\label{eq:Emeasure}
	E=X_3^2+X_4^2+X_5^2+X_6^2+X_7^2+X_8^2=1-X_1^2-X_2^2-X_9^2
\end{equation}

\subsection{\label{sec:sbusec23}Relation with two-tangle and three-tangle}

The quantity (\ref{eq:Emeasure}) is directly related to an entanglement measures known as three-tangle and two-tangle \cite{MIY, COFF}. These measures depends of the hyperdeterminant of the  tensor with coefficients $\left(a_{ijk}\right)$ that define the state $\left|\Psi\right\rangle$ in (\ref{eq:threequbits}). For a tensor $A$ with components $a_{ijk}$, this hyperdeterminant is defined as \cite{BORS}
\begin{equation}
\label{eq:HDet}
\text{det}(A)=\frac{1}{2}\epsilon^{il}\epsilon^{jm}c_{ij}c_{lm}\quad c_{kn}=\frac{1}{2}\epsilon^{il}\epsilon^{jm}a_{ijk}a_{lmn}
\end{equation}
where $\epsilon$ in (\ref{eq:HDet}) is the Levi-Civita symbol. The  three-tangle measure gives information about entanglement  between all three qubits $(A, B, C)$ and is given by
\begin{equation}
\label{eq:3tangle}
\tau_{ABC}= 4\text{det}(A).
\end{equation}
The two-tangle measure gives the information about entanglement between one subsystem (e.g. $A$) and the other two subsystems  ($BC$):
\begin{eqnarray}
\label{eq:2tangle}
\tau_{A(BC)}= 4\text{det}(\rho_A)&&\tau_{B(CA)}= 4\text{det}(\rho_B)\nonumber\\
\tau_{C(AB)}&=& 4\text{det}(\rho_C). 
\end{eqnarray}
Explicitly,
\begin{widetext}
\begin{equation}
\label{eq:2tangle2}
\tau_{A(BC)}= 4\text{det}
\left(
\begin{array}{ll}
 a_{000}^2+a_{001}^2+a_{010}^2+a_{011}^2 & a_{010} a_{110}+a_{011} a_{111}+a_{000} a_{100}+a_{001} a_{101} \\
 a_{010} a_{110}+a_{011} a_{111}+a_{000} a_{100}+a_{001} a_{101} & a_{100}^2+a_{101}^2+a_{110}^2+a_{111}^2,
\end{array}
\right)
\end{equation}
\end{widetext}
expression (\ref{eq:Emeasure}) matches perfectly with $\tau_{A(BC)}$ in (\ref{eq:2tangle2}), that coincides with the fact that  expression (\ref{eq:Emeasure}) is obtained considering that the state $\left|\Psi\right\rangle$ in (\ref{eq:threequbits}) is bi-separable. For bi-separable states, $\tau_{A(BC)}\neq 0$, but $\tau_{ABC}=0$ as in the $W$-states where all $2$-tangles do not vanish but $3$-tangle is still zero; unlike the fully separable states where all $2$-tangles and $3$-tangle vanish. Finally, when the state is maximally entangled, $2$-tangles and the $3$-tangle are non-zero as $GHZ$-state.

\section{\label{sec:sec3}Geometric representation of four qubits}

This section emphasizes in a possible map (as the Hopf fibration for two and three qubit states) that helps us to reduce the Hilbert space of four qubits to a lower space and gives an entanglement interpretation of this geometric illustration. For this, we continue the iteration of  the Cayley-Dickson construction and we use sedenions. 	The real, complex, quaternion and octonion numbers define a \emph{division algebra}. The octonions are the biggest division algebra since in sedenions the division property is lost, this is because they have zero divisors, that means that two non-zero sedenions can be multiplied to obtain a zero result.\\

Mathematically, the Hopf fibration in this case does not exist, this tells us that the systems of one, two and three qubits are the only systems that accept a Hopf fibration interpretation \cite{CHEN}. To find a possible map that reduces the original Hilbert space of four qubits, as the Hopf fibrations does with the previous cases is our goal. The idea is to introduce a new map like a Hopf fibration with restrictions and study some specific examples as $GHZ$ and $W$ states.\\

The general pure state of four qubits is given by
\begin{eqnarray}
\label{eq:fourqubits}
&&\left|\Psi\right\rangle=a_{ABCD}\left|ABCD\right\rangle=\nonumber\\	&&a_{0000}\left|0000\right\rangle+a_{0001}\left|0001\right\rangle+a_{0010}\left|0010\right\rangle+a_{0100}\left|0100\right\rangle+\nonumber\\
&&a_{1000}\left|1000\right\rangle+a_{0011}\left|0011\right\rangle+a_{0110}\left|0110\right\rangle+a_{1100}\left|1100\right\rangle+\nonumber\\	&&a_{0101}\left|0101\right\rangle+a_{1001}\left|1001\right\rangle+a_{1010}\left|1010\right\rangle+a_{0111}\left|0111\right\rangle+\nonumber\\
&&a_{1110}\left|1110\right\rangle+a_{1011}\left|1011\right\rangle+a_{1101}\left|1101\right\rangle+a_{1111}\left|1111\right\rangle\nonumber\\
&&
\end{eqnarray}
the normalization condition in this case reads		
\begin{eqnarray}
\label{eq:normalization3}	&&\left|a_{0000}\right|^2+\left|a_{0001}\right|^2+\left|a_{0010}\right|^2+\left|a_{0100}\right|^2+\left|a_{1000}\right|^2+\nonumber\\
&&\left|a_{0011}\right|^2+\left|a_{0110}\right|^2+\left|a_{1100}\right|^2+\left|a_{0101}\right|^2+\left|a_{1001}\right|^2+\nonumber\\
&&\left|a_{1010}\right|^2+\left|a_{0111}\right|^2+\left|a_{1110}\right|^2+\left|a_{1011}\right|^2+\left|a_{1101}\right|^2+\nonumber\\
&&\left|a_{1111}\right|^2=1.	
\end{eqnarray}
This condition identifies the Hilbert space of four qubits to a $\bm{S}^{31}\in\mathbb{R}^{32}$. It is important to understand that the entanglement of four qubits is not so simple to explain as we did before with three and two qubits. The different ways that four qubits can be entangled are more complicated that the cases studied in the previous sections. Furthermore,  four qubits can be entangled in nine different ways \cite{NINE}. Our interest is emphasized in the case when one qubit (e.g. qubit $A$) is separated of the other three, it is important to note that this case is independent of the qubit choice, it can be any of them. In this situation the pure qubit state can be written as
\begin{eqnarray}
\label{eq:fourseparation} 	
\left|\Psi\right\rangle&=&\left(a\left|0\right\rangle+b\left|1\right\rangle\right)\otimes\nonumber\\	&&(c\left|000\right\rangle+d\left|001\right\rangle+e\left|010\right\rangle+f\left|011\right\rangle+\nonumber\\
&&g\left|100\right\rangle+h\left|101\right\rangle+m\left|110\right\rangle+n\left|111\right\rangle)\nonumber\\
&\Rightarrow&\left(\left|a\right|^2+\left|b\right|^2\right)\nonumber\\
&&(\left|c\right|^2+\left|d\right|^2+\left|e\right|^2+\left|f\right|^2+\nonumber\\
&&\left|g\right|^2+\left|h\right|^2+\left|m\right|^2+\left|n\right|^2)=1.
\end{eqnarray}
With (\ref{eq:fourseparation}) it is possible to find the separability conditions, some of them are:
\begin{eqnarray}
\label{eq:fourconditions2} 
a_{0010}a_{1000}=a_{1010}a_{0000}&& a_{0110}a_{1100}=a_{1110}a_{0100} \nonumber\\
a_{0000}a_{1011}&=&a_{0011}a_{1000}...	
\end{eqnarray}
For the four qubits case, we expect that the information about the separated qubit and its entanglement with the other three qubits is  stored in the base $\bm{S}^{16}$. Also, we anticipate that the information of the other three qubits  is in the fibre $\bm{S}^{15}$ and the corresponding entanglement analysis is possible with the fibrations studied. Similar to the previous cases, the possible map should be
\begin{subequations}
\label{eq:wholemap3}
\begin{eqnarray}
h''_1:\qquad\bm{S}^{31} &\rightarrow& \mathbb{R}^{16}+\left\{\infty\right\}\nonumber\\
\mathbb{S}\otimes \mathbb{S} &\rightarrow& \mathbb{S}\cup\left\{\infty\right\} \approx \bm{S}^{16}\nonumber\\
\left(s_1, s_2\right)&\rightarrow& h''_1= s_1s_2^{-1}	\qquad s_1, s_2 \in \mathbb{S}
\label{subeq:h111}
\end{eqnarray}
\begin{eqnarray}
h''_2: \qquad\mathbb{R}^{16}+\left\{\infty\right\} &\rightarrow& \bm{S}^{16}\nonumber\\
\mathbb{S}\cup\left\{\infty\right\} &\rightarrow& \bm{S}^{16}
\label{subeq:h222}
\end{eqnarray}
\begin{equation}
\label{subeq:h111h222}
h''_2\circ h''_1(o_1,o_2)=X_i=\left\langle \sigma_i\right\rangle\qquad i={1,...,17}
\end{equation}
\end{subequations}
where $\left\langle \sigma_i\right\rangle$ in (\ref{subeq:h111h222}) are the Pauli matrices ti sedenionic space
\begin{eqnarray}
\label{eq:Spauli}
\sigma_1&=&\left(
\begin{array}{ll}
0 & 1 \\
1 & 0
\end{array}
\right)\nonumber\\
\sigma_{2,...,16}&=&\left(
\begin{array}{ll}
0                   & i_{1,...,15} \\
-i_{1,...,15}  & 0
\end{array}
\right)\nonumber\\
\sigma_{17}&=&\left(
\begin{array}{ll}
1    & 0 \\
0    & -1.
\end{array}
\right)
\end{eqnarray}
The sedenions in (\ref{subeq:h111}) are defined from the complex coefficients of the state $\left|\Psi\right\rangle$ in (\ref{eq:fourqubits}). This map (\ref{eq:wholemap3}) is not always possible since the division property is lost for sedenions, this gives us an impossibility of a broad map for a general four qubit pure state, and the fibre of the map is not clear since the multiplication of sedenions is neither commutative, associative nor alternative. Nevertheless, it is possible to construct two sedenions in which case the multiplication is not null. For the Cayley-Dickson construction it is indispensable to define eight quaternions, four octonions and finally two sedenions. Our construction is given by
\begin{eqnarray}	
\label{eq:sedenions}
q_1&=&a_{0000}+a_{0001}\bm i_2 \quad q_2=a_{0010}+a_{0011}\bm i_2\nonumber\\
q_3&=&a_{0100}+a_{0101}\bm i_2 \quad q_4=a_{0110}+a_{0111}\bm i_2\nonumber\\
q_5&=&a_{1000}+a_{1001}\bm i_2 \quad q_6=a_{1010}+a_{1011}\bm i_2\nonumber\\
q_7&=&a_{1100}+a_{1101}\bm i_2 \quad q_8=a_{1110}+a_{1111}\bm i_2\nonumber\\
&&\nonumber\\
o_1&=&q_1+q_2 \bm i_4 \quad o_2=q_3+\bar{q}_4 \bm i_4\nonumber\\
o_3&=&q_5+q_6 \bm i_4 \quad o_4=q_7+\bar{q}_8 \bm i_4\nonumber\\
&&\nonumber\\
s_1&=&o_1+o_2 \bm i_8 \quad s_2=o_3+o_4 \bm i_8.
\end{eqnarray} 
And the possible coordinates that define the $\bm{S}^{16}$ are
\begin{eqnarray}
\label{eq:finalcoordinatesSED}	
	X_1&=&s_1\overline s_2+s_2\overline s_1 \nonumber\\
	X_2&=&\text{Re}\left[\bm{i}_1(s_1\overline s_2-s_2\overline s_1)\right]\nonumber\\
	&\vdots&\nonumber\\
	X_{16}&=&\text{Re}\left[\bm{i}_{15}(s_1\overline s_2-s_2\overline s_1)\right]\nonumber\\
	X_{17}&=&s_1\overline s_1-s_2\overline s_2  
\end{eqnarray} 
Analogous to the three and two qubit cases we expect that this map (\ref{eq:wholemap3}) is entanglement sensitive. Moreover, in the case when one qubit is separated from the other three, only three of the last coordinates in (\ref{eq:finalcoordinatesSED}) are not null. The first part of the whole map (\ref{subeq:h111}) is now
\begin{eqnarray}
\label{eq:finalh1SED}		h_1''(s_1,s_2)&=&s_1s_2^{-1}=\frac{C_1+C_2\bm{i}_4+C_3\bm{i}_8+C_4\bm{i}_{12}}{\left|q_5\right|^2+\left|q_6\right|^2+\left|q_7\right|^2+\left|q_8\right|^2}\nonumber\\
&&\\
C_1&=&q_1\bar q_5+\bar q_6 q_2+\bar q_7 q_3+q_4 \bar q_8\nonumber\\
C_2&=&q_2q_5-q_6q_1+\overline{(q_4q_7-q_8q_3)}\nonumber\\
C_3&=&q_3q_5-q_7q_1+\overline{(q_3q_8-q_6q_4)}\nonumber\\
C_4&=&q_2q_7-q_6q_3+\overline{(q_8q_1-q_4q_5)}
\end{eqnarray}
In the generic case, $h''_1$ is a sedenion. Nevertheless with some of the conditions (\ref{eq:fourconditions2}), in which the four qubits are separable as  one-qubit $\otimes$ three-qubits, the numbers $K_2, K_3$ and $K_4$ are zero, evenmore $K_1$ is reduced to a complex number, and $h''_1$ maps into the subspace of pure complex numbers $\mathbb{C}\cup\infty$ in the sedenionic field $\mathbb{S}\cup\infty$ as in the last case, and it is proved that \emph{this map is also entanglement sensitive}.\\
\begin{equation}
\label{eq:h1separableSED}
h''_1(s_1,s_2)|_{\text{separable}}=\frac{C_1}{\left|q_5\right|^2+\left|q_6\right|^2+\left|q_7\right|^2+\left|q_8\right|^2}\in \mathbb{C}\cup\infty
\end{equation}
In the separable case, $K_1$ is always different from zero, so the multiplication of sedenions defined in (\ref{eq:sedenions}) is always different from zero.  If the state (\ref{eq:fourqubits}) is entangled (specifically one qubit with the other three), then $K_2$, $K_3$ and $K_4$ are not null and the map defined (\ref{eq:wholemap3}) is always possible with the sedenions defined as (\ref{eq:sedenions}) and it is useful for the study of the quantum correlation of one qubit with the other three. The importance of  (\ref{eq:sedenions}) lies in the fact that the multiplication of these sedenions is not null.\\

Analogous to the last case, since the coordinates $X_3,...,X_{17}$ are zero for separable states, and non-null for entangled states, it is possible to characterize the degree of entanglement of one qubit with the other three qubits using these coordinates, such that   
\begin{equation}
\label{eq:NEWmedida}
E=X_3^2+X_4^2+...+X_{16}^2=1-X_1^2-X_2^2-X_{17}^2.
\end{equation}
The next step with this entanglement measure is to compare it with other proposed measures \cite{MIY, YAN, DAN, BOR} and give the value of $E$ for  some known states. The first important thing is to note that  $E$ in (\ref{eq:NEWmedida}) is different from the three-tangle \cite{WONG} defined in this case as $\tau_{A(BCD)}=4 \text{ det}(\rho_{A})$ where $\rho_{A}$ is the reduced matrix defined as $\rho_A=\text{Tr}_{BCD}\left|\Psi_{ABCD}\right\rangle\left\langle \Psi_{ABCD}\right|$, like in the case of three qubits.\\

The entanglement degree for some known states as $GHZ$ and $W$ states is

\begin{eqnarray}
\label{eq:EXfour}
&&\left|GHZ\right\rangle=\frac{\left|0000\right\rangle+\left|1111\right\rangle}{\sqrt{2}}, \quad E = 1\nonumber\\	&&\left|W\right\rangle_0=\frac{1}{2}\left(\left|1000\right\rangle+\left|0100\right\rangle+\left|0010\right\rangle+\left|0001\right\rangle\right),\quad  E = \frac{1}{2}\nonumber\\	&&\left|W\right\rangle_1=\frac{1}{2}\left(\left|0111\right\rangle+\left|1011\right\rangle+\left|1101\right\rangle+\left|1110\right\rangle\right),\quad E = \frac{3}{4}\nonumber\\
&&\left|\Phi\right\rangle_1=\frac{1}{\sqrt{6}}\left(\sqrt{2}\left|1111\right\rangle+\left|1000\right\rangle+\left|0100\right\rangle+\left|0010\right\rangle+\left|0001\right\rangle\right),\nonumber\\
&&\qquad E = \frac{8}{9}\nonumber\\
&&\left|\Phi\right\rangle_2=\frac{1}{\sqrt{2\sqrt{10}}}(3\left|0000\right\rangle+3\left|1111\right\rangle-\left|0011\right\rangle-\left|1100\right\rangle\nonumber\\
&&\qquad+3\left|0101\right\rangle+3\left|1010\right\rangle-\left|0110\right\rangle-\left|1001\right\rangle),  \nonumber\\
&&\qquad E = 0.6625
\end{eqnarray}
The measure of entanglement of some states as $GHZ$-state has coherent results. For $GHZ$ state $E$ has the maximum value, for $W$ states $E$ is different from zero and less that one, it show us that these states are entangled as we expected. The  state $\left|\Phi\right\rangle_1$ of  (\ref{eq:EXfour}) has the same entanglement degree given by A. Osterloh and J. Siewet in \cite{AND}  who define an entanglement monotone from antilinear operators, but for example their results for $W$ states are zero, That differs from our results.\\

For maximally entangled states (MES) the coordinates $X_3,...,X_{16}$ have the maximum possible value since $X_1=X_2=X_{17}=0$, then the corresponding normalization conditions in this case is,

\begin{equation}
\label{eq:MES}
X_3^2+X_4^2+...+X_{16}^2=1 \quad \text{MES} \in \bm{S}^{13}.
\end{equation}
	
The MES span a thirteen-dimensional sphere. Since the information of one qubit is contained in  $\bm{S}^{16}$ and the information of the other three qubits is in  $\bm{S}^{15}$, it is possible to take a subspace in  $\bm{S}^{16}$ defined from the coordinates $\left(X_1, X_2, X_{17}\right)$ such that all the four qubits states are mapped onto a unit Ball $B^3$. The separable states are mapped onto the boundary and the entangled states inside the ball; the center of the ball represent the MES. The states with the same degree of entanglement are in the same concentric shells, it means that the radius of this ball is directly connected with the degree of the entanglement of the four qubit state.

\section{\label{sec:sec4}Conclussion and Discussion}

The Hopf fibration is a good mathematical object that helps us to visualize the multiple-qubit systems. Furthermore, this point of view is entanglement sensitive since the base of the fibration is reduced to a lower dimensional space when the  respective separability conditions are fulfilled. The entanglement measures defined with the Hopf maps match exactly with known and well defined entanglement measures such as concurrence and tangles for two and three qubit cases  respectively. A new map is proposed for four qubit case using sedenions and an entanglement measure is introduced to describe the entanglement of the four-qubit state and the possibility of it being separable as a one-qubit $\otimes$ three-qubits. The generalization to n-qubit systems is nontrivial but the fruitfulness of this geometric interpretation gives  important information about a possible correspondence for entanglement of higher qubit states which is a matter of research in current investigations e.g. \cite{BORS, YAN}.


\end{document}